\documentclass[prb,12pt,preprint,a4paper,groupedaddress,endfloats]{revtex4-1}

\usepackage{times}
\usepackage{mathptmx}

\usepackage{amsmath,amssymb,latexsym}
\usepackage{bm}
\usepackage{graphicx}

\newcommand{\p}{\partial}

\newcommand{\spar}{\parallel}
\newcommand{\vek}[1]{\ensuremath{\mathbf{#1}}}

\renewcommand{\tensor}[1]{\mathsf{#1}}

\newcommand{\ave}[1]{\left<#1\right>}
\newcommand{\abs}[1]{{\left|#1\right|}}
\newcommand{\order}[1]{\mathcal{O}(#1)}

\newcommand{\rmd}{\mathrm{d}}

\newcommand{\bxi}{\bm{\xi}}

\newcommand{\bmu}{\bm{\mu}}

\newcommand{\blambda}{\bm{\lambda}}

\newcommand{\ddt}[1]{\frac{\rmd#1}{\rmd t}}
\newcommand{\ddT}[1]{\frac{\rmd#1}{\rmd T}}

\newcommand{\pdX}[1]{\frac{\partial#1}{\partial\vek{X}}}
\newcommand{\pdt}[1]{\frac{\partial#1}{\partial t}}
\newcommand{\pdT}[1]{\frac{\partial#1}{\partial T}}

\newcommand{\pdtau}[1]{\frac{\partial#1}{\partial\tau}}

\newcommand{\Eqref}[1]{Eq.~(\ref{#1})}
\newcommand{\Eqsref}[1]{Eqs.~(\ref{#1})}
\newcommand{\Figref}[1]{Fig.~\ref{#1}}

\newcommand{\Secref}[1]{Sec.~\ref{#1}}

\begin{document}

\title{Charged particle motion in weakly varying electromagnetic fields:\\ A multi-scale approach}

\author{D.~S.~Sortland}
\author{O.~E.~Garcia}
\email{Corresponding author: odd.erik.garcia@uit.no}
\affiliation{Department of Physics and Technology, UiT The Arctic University of Norway, N-9019 Troms{\o}, Norway}

\date{\today}

\begin{abstract}
The motion of charged particles in weakly varying electromagnetic fields is described using a perturbation method. This provides a systematic and physically transparent description of the particle motion on fast and slow spatio-temporal scales, associated with gyration and drift motions, respectively. A detailed discussion is given of the guiding center concept and the non-inertial frame of reference. An algebraic expression is obtained for the drift motion across the magnetic field, while a differential equation describes the particle motion along the field. The fictitious forces and associated energy transfer between gyration and drift motion parallel and perpendicular to the magnetic field are described. The relation between conservation of magnetic moment and angular momentum is elucidated.
\end{abstract}

\maketitle

\section{Introduction}\label{intro}

The knowledge of charged particle motion in weakly varying electromagnetic fields is fundamental for our understanding of collective phenomena in magnetized plasmas. Accordingly, most textbooks in plasma physics have an introductory part describing single particle motion in prescribed fields.\cite{balescu,bellan,chen,friedberg,gr,hm,hw,parks,pecseli}

For constant and uniform electromagnetic fields there is an exact solution of the equations of motion which comprises three parts: particle gyration around a guiding center that is accelerated along the magnetic field and a constant drift perpendicular to both the electric and magnetic fields. In the case of fields with slow temporal and weak spatial variations, exact solutions of the equations of motion can generally not be obtained.

Approximate solutions can be found by use of a perturbation theory using the well-known exact solution for constant and uniform fields as the lowest order solution. We present a systematic and physically transparent solution of this problem using a perturbation method, separating the scales associated with particle gyration and variations of the electromagnetic fields.

In addition to deriving an approximate solution for the guiding center motion, we present details about the choice of a magnetic field based coordinate system, the non-inertial frame of reference associated with the guiding center system, the corresponding fictitious forces, multi-scale expansion of dependent variables, and the conservation of energy and magnetic moment.

The outline of this tutorial is as follows. In \Secref{constant} we describe the exact analytical solution for charged particle motion in constant and uniform fields, introducing the guiding center concept. In \Secref{varying} we describe the particle motion in the case of weakly varying fields, including a discussion of the perturbation method used. The conservation of energy, angular momentum and magnetic moment is discussed in detail. A summary of the results are presented in \Secref{sec:concl}.

\section{Motion in constant and uniform fields}\label{constant}

In this section we describe the exact solution of the equations of motion for a particle moving in constant and uniform electric and magnetic fields, emphasizing the magnetic field based coordinate representation, the role of the initial conditions for the particle motion, the non-inertial frame of reference and energy conservation.

\subsection{Orthonormal magnetic coordinates}\label{b-unit}

The presence of a magnetic field leads to fundamentally different particle motion in the directions along and perpendicular to the field lines. In order to describe the particle motion, it is advantageous to introduce a Cartesian coordinate system where $\vek{e}_3=\vek{B}/B$ is the unit vector along the magnetic field, and $\vek{e}_1$ and $\vek{e}_2$ are unit vectors spanning the plane perpendicular to the magnetic field, defined by
\begin{equation}\label{GS}
\vek{e}_1 = \frac{\left(\vek{C}\times\vek{B}\right)\times\vek{B}}{B\left|\vek{C}\times\vek{B}\right|} ,
\qquad
\vek{e}_2 = \frac{\vek{C}\times\vek{B}}{\left|\vek{C}\times\vek{B}\right|} ,
\end{equation}
where $\vek{C}$ is an arbitary but constant vector with a non-vanishing component perpendicular to $\vek{B}$. In the following, we will also use the conventional notation $\vek{b}=\vek{e}_3$ as the unit vector along the magnetic field. The unit vectors $\vek{e}_1$, $\vek{e}_2$ and $\vek{e}_3$ form a right-handed orthonormal basis, $\left(\vek{e}_1\times\vek{e}_2\right)\cdot\vek{e}_3=1$. As a consequence of the anisotropy imposed by the magnetic field, we will in the following decompose any vector $\vek{A}=A_1\vek{e}_1+A_2\vek{e}_2+A_3\vek{e}_3$ into a component along the magnetic field, $\vek{A}_\spar=(\vek{b}\cdot\vek{A})\vek{b}=A_3\vek{e}_3$, and its components perpendicular to the magnetic field, $\vek{A}_\perp=\vek{b}\times(\vek{A}\times\vek{b})=A_1\vek{e}_1+A_2\vek{e}_2$.

\subsection{Particle motion} \label{motion}

The equations of motion in an inertial frame of reference for a particle with mass $m$, electric charge $q$, position $\vek{x}$ and velocity $\vek{v}$ in an electric field $\vek{E}$ and magnetic field $\vek{B}$ are given by
\begin{subequations} \label{eom}
\begin{align} 
\ddt{\vek{x}} & = \vek{v} , \label{eomA}
\\
m\,\ddt{\vek{v}} & = q( \vek{E} + \vek{v}\times\vek{B} ) , \label{eomB}
\end{align}
\end{subequations}
with initial conditions $\vek{x}(t=0)=\vek{x}_*$ and $\vek{v}(t=0)=\vek{v}_*$. For constant and uniform electromagnetic fields, the general solution for the particle velocity can be written as
\begin{equation} \label{vexact}
\vek{v} = \vek{u} + \vek{U} = \vek{u} + \vek{U}_\spar + \vek{U}_E .
\end{equation}
Here $\vek{u}$ describes gyration motion of the particle in the plane perpendicular to the magnetic field, governed by the equation
\begin{equation} \label{quB}
\ddt{\vek{u}} = \omega\vek{u}\times\vek{b} ,
\end{equation}
where $\omega=qB/m$ is the gyration frequency. Taking the time derivative of \Eqref{quB} shows that the $\vek{B}$-perpendicular components of the velocity vector satisfy the equation for harmonic oscillations,
\begin{equation}
\frac{\rmd^2\vek{u}}{\rmd t^2} - \omega^2\vek{u} = \vek{0} .
\end{equation}
The solution of \Eqref{quB} can be written as
\begin{equation} \label{uexact}
\vek{u}(t) = u\left[ \vek{e}_1 \cos(\omega t+\theta) + \vek{e}_2 \sin(\omega t+\theta) \right] ,
\end{equation}
where $\theta$ is the initial gyro-phase. The gyration speed $u$ and the initial gyro-phase $\theta$ are given by the initial value conditions, discussed further in the following subsection. Note that $\vek{u}$ does not have any component along the magnetic field, that is, $\vek{u}\cdot\vek{b}=0$, such that $\vek{u}$ only describes the gyration motion.

The velocity vector $\vek{U}_\spar$ describes the particle motion along the magnetic field and satisfies the equation
\begin{equation}\label{Uspardyn}
m\,\ddt{U_\spar} = qE_\spar ,
\end{equation}
with the solution given by
\begin{equation} \label{Uspar}
U_{\spar}(t) = v_{*\spar} + \frac{qE_\spar t}{m} ,
\end{equation}
which describes uniform acceleration by the $\vek{B}$-parallel component of the electric field. The last term in \Eqref{vexact} is the \emph{electric drift},
\begin{equation} \label{ExB}
\vek{U}_E = \frac{\vek{E}\times\vek{B}}{B^2} ,
\end{equation}
which describes a constant motion of the particle perpendicular to both the electric and the magnetic fields. This drift is independent of the particle mass, charge and gyration velocity. The electric drift is thus the same for all charged particles, and represents bulk motion of a plasma.

\begin{figure}[t]
\centering
\includegraphics[width=8cm]{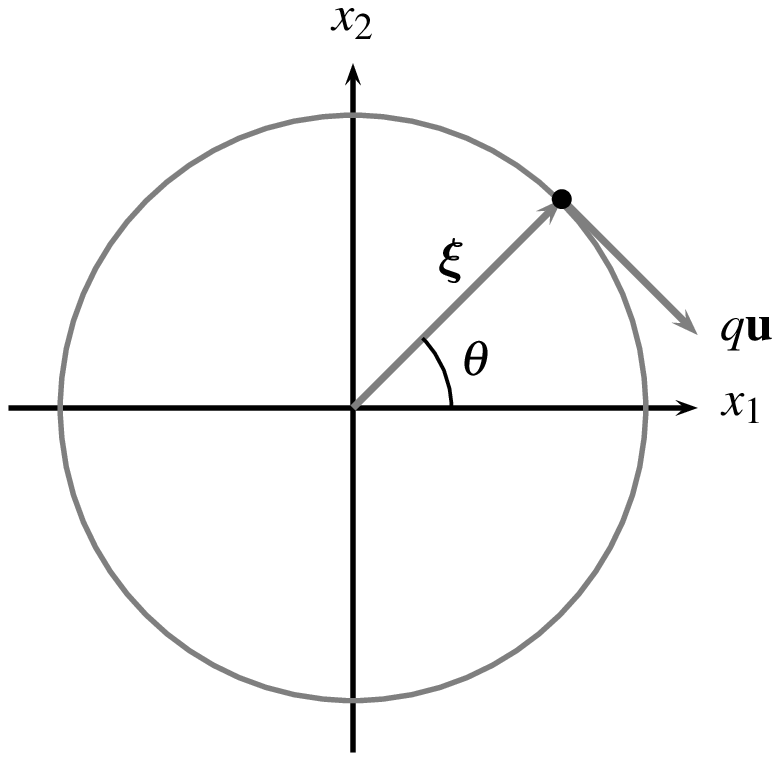}
\caption{Illustration of gyration orbit for a particle with charge $q$ in the $(x_1,x_2)$-plane perpendicular to the magnetic field. The particle position is shown at the initial time $t=0$, which defines the initial gyro-phase $\theta$. The initial gyration radius $\bxi$ and velocity $\vek{u}$ are shown. The magnetic field points out of the paper plane.}
\label{fig:gyro}
\end{figure}

The solution for the particle position is obtained by straight forward integration of the velocity vector $\vek{v}$, and can be written as
\begin{equation} \label{xxiX}
\vek{x} = \bxi + \vek{X} ,
\end{equation}
where we have defined the gyration vector $\bxi$ and the instantaneous center of gyration $\vek{X}$ by
\begin{subequations} \label{xexact}
\begin{align}
\bxi(t) & = \xi \left[ \vek{e}_1 \sin(\omega t+\theta) - \vek{e}_2 \cos(\omega t+\theta) \right] ,  \label{xexacta}
\\
\vek{X}(t) & = \vek{X}_* + \left(\vek{U}_\spar+ \vek{U}_E\right)t . \label{xexactb}
\end{align}
\end{subequations}
Here the gyration radius is given by $\xi=u/\omega=mu/qB$ and $\vek{X}_*=\vek{x}_*-\bxi(t=0)$ is the initial center of gyration. The particle velocities are related to the positions by $\vek{u}=\rmd\bxi/\rmd t$ and $\vek{U}=\rmd\vek{X}/\rmd t$. It follows that the gyration velocity can be written in coordinate-free form in terms of gyration frequency and radius as
\begin{equation} \label{uoxb}
\vek{u} = \omega\bxi\times\vek{b} ,
\end{equation}
a relation that will be used repeatedly in the following section. The gyration motion is illustrated in \Figref{fig:gyro}.

From the above solution it follows that the particle position vector satisfies the equation of a circle,
\begin{equation}
( \vek{x} - \vek{X} )^2 = \xi^2 .
\end{equation}
The instantaneous center of gyration $\vek{X}(t)=\vek{X}_*+\vek{U}t$, in the following referred to as the \emph{guiding center}, moves along the along the magnetic field with velocity $\vek{U}_\spar(t)$ and perpendicular to the field with the electric drift $\vek{U}_E$.

\subsection{Magnetic moment}\label{moment}

It should be noted that the sign of $\omega$ and $\xi$ depend on the particle charge $q$. It follows from \Eqref{xexacta} that both negatively and positively charged particles will gyrate such as to reduce the magnetic field inside the gyration orbit. The gyration motion thus implies that an electrically charged particle moving in a magnetic field constitutes a magnetic dipole. The magnetic dipole moment is given by the average current $q\omega/2\pi$ times the area $\pi\xi^2$,
\begin{equation}
\bmu = - \frac{mu^2}{2B}\,\vek{b} .
\end{equation}
Accordingly, a magnetized plasma is intrinsically diamagnetic. The potential energy of a dipole in a magnetic field is given by $P=-\bmu\cdot\vek{B}$ and there is a net force on the dipole given by $-\nabla P$. In the case of constant and uniform fields the magnetic moment is evidently a constant of motion and this force vanishes. In the next section this force will be derived from the equation of motion for the case of weakly varying fields, and the magnetic moment will be shown to be an approximate constant of motion.

\subsection{Initial conditions}\label{initial}

The gyration speed $u$ and the initial gyro-phase $\theta$ are related to the initial particle velocity and the electromagnetic fields by the relation
\begin{equation}\label{initialvelocityperp}
\vek{v}_{*\perp}  =u\left[ \vek{e}_1 \cos\theta + \vek{e}_2 \sin\theta \right] +  \vek{U}_E ,
\end{equation}
which follows directly from \Eqref{vexact}. From this it follows that the gyration speed can be written as
\begin{equation}\label{gyrationvelocity}
u=\sqrt{\left(\vek{v}_{*\perp}-\vek{U}_E\right)^{2}} ,
\end{equation}
which is the difference between the initial $\vek{B}$-perpendicular particle velocity and the electric drift. For the special case of a particle initially at rest, the speed of gyration is given by the electric drift.

Taking the projection of \Eqref{initialvelocityperp} on $\vek{e}_1$ and $\vek{e}_2$ and dividing the resulting equations gives the initial gyro-phase dependence on the initial particle velocity,
\begin{equation}
\tan\theta = \frac{\vek{e}_2\cdot(\vek{v}_{*\perp}-\vek{U}_E)}{\vek{e}_1\cdot(\vek{v}_{*\perp}-\vek{U}_E)} .
\end{equation}
Given the electromagnetic fields and the initial conditions $\vek{x}(t=0)=\vek{x}_*$ and $\vek{v}(t=0)=\vek{v}_*$, the initial gyro-phase, gyration radius and speed is readily found from the above relations. Rather than the initial $\vek{B}$-perpendicular particle velocity components $\vek{v}_{*\perp}$ we may alternatively give $\theta$ and $\xi$ or $u$ as initial conditions.

\subsection{Frame of reference}\label{frame}

The results above show that in the case of constant and uniform magnetic fields, the particle motion can be described as uniform gyration with position vector $\bxi=\xi_1\vek{e}_1+\xi_2\vek{e}_2$ in the guiding center frame of reference, and a drift of the guiding center with position vector $\vek{X}$ in the inertial frame. The particle is accelerated along the magnetic field, and the guiding center frame of reference is therefore not an inertial frame. The temporal change of the particle position vector in the inertial frame, $\vek{x}=\vek{X}+\bxi$, is given by
\begin{equation} \label{position}
\ddt{\vek{x}} = \ddt{\vek{X}} + \sum_{j=1}^3 \ddt{\xi_j}\,\vek{e}_j + \sum_{j=1}^3 \xi_j\,\ddt{\vek{e}_j} .
\end{equation}
Here the first term on the right hand side describes the velocity of the guiding center with respect to the inertial frame, $\vek{U}=\rmd\vek{X}/\rmd t$, while the second term is the particle velocity as observed in the guiding center frame of reference,
\begin{equation}
\vek{u} = \sum_{j=1}^3 u_j\vek{e}_j = \sum_{j=1}^3 \ddt{\xi_j}\,\vek{e}_j .
\end{equation}
When the magnetic field is not constant, derivatives of the unit vectors $\vek{e}_j$ for $j=1,2,3$ correspond to rotation of the guiding center system, so the last term in \Eqref{position} results from an apparent change of the particle velocity due to rotation of the coordinate axes. Indeed, a point observed in the guiding center frame of reference has a rotational component of the velocity that increases with the distance $\xi$ from the origin. Of course, the last term in \Eqref{position} vanishes for a constant and uniform magnetic field since in this case $\rmd\vek{e}_j/\rmd t=\vek{0}$ for $j=1,2,3$.

Taking a time derivative of \Eqref{position} provides an equation for the acceleration in the non-inertial guiding center frame of reference,
\begin{equation}
\ddt{\vek{v}} = \ddt{\vek{U}} + \sum_{j=1}^3 \ddt{u_j}\,\vek{e}_j + \sum_{j=1}^3 2u_j\,\ddt{\vek{e}_j} + \sum_{j=1}^3 \xi_j\,\frac{\rmd^2\vek{e}_j}{\rmd t^2} ,
\end{equation}
where $\vek{v}=\rmd\vek{x}/\rmd t$. The term on the left hand side describes the acceleration of the particle in the inertial frame of reference. The first term on the right hand side is the acceleration of the guiding center as observed in the inertial frame. The second term on the right hand side due to the acceleration of the particle as observed in the guiding center system, while the two last terms describe an apparent acceleration due to rotation of the guiding center frame. In the guiding center frame of reference, the particle is thus subject to \emph{fictitious forces},
\begin{equation}
\sum_{j=1}^3 m\,\ddt{u_j}\,\vek{e}_j = m\,\ddt{\vek{v}} + \vek{F}_\text{fictitious} ,
 \end{equation}
where the fictitious forces comprise one part due to the relative motion between the inertial and guiding center systems, and two parts due to rotation of the guiding center system,
\begin{equation}
\vek{F}_\text{fictitious} = - m\,\ddt{\vek{U}} - 2m \sum_{j=1}^3 u_j\,\ddt{\vek{e}_j} - m  \sum_{j=1}^3 \xi_j\,\frac{\rmd^2\vek{e}_j}{\rmd t^2} .
\end{equation}
Using the equation of motion \eqref{eom} with $\vek{v}=\vek{u}+\vek{U}$ in the Lorentz force, the equation of motion in the guiding center frame of reference can be written as
\begin{equation}
\sum_{j=1}^3 m\,\ddt{u_j}\,\vek{e}_j = q\vek{E} + q\vek{u}\times\vek{B} + q\vek{U}\times\vek{B} - m\,\ddt{\vek{U}} - 2m \sum_{j=1}^3 u_j\,\ddt{\vek{e}_j} - m  \sum_{j=1}^3 \xi_j\,\frac{\rmd^2\vek{e}_j}{\rmd t^2} .
\end{equation}
In this frame, the particle motion is just uniform gyration, described by \Eqref{quB},
\begin{equation} \label{ugyr}
\sum_{j=1}^3 m\,\ddt{u_j}\,\vek{e}_j = q\vek{u}\times\vek{B} .
\end{equation}
We thus obtain the equation describing the guiding center motion,
\begin{equation} \label{gceomf}
m\,\ddt{\vek{U}} = q\vek{E} + q\vek{U}\times\vek{B} - 2m \sum_{j=1}^3 u_j\,\ddt{\vek{e}_j} - m  \sum_{j=1}^3 \xi_j\,\frac{\rmd^2\vek{e}_j}{\rmd t^2} .
\end{equation}
In the case of a constant and uniform magnetic field, the last two terms on the right hand side vanish and we obtain
\begin{equation} \label{gceom}
m\,\ddt{\vek{U}} = q\vek{E} + q\vek{U}\times\vek{B} ,
\end{equation}
where the term on the left hand side is the fictitious force. This is exactly the problem solved in \Secref{motion}, describing the motion of the guiding center. In the guiding center frame of reference, the particle is not subject to any electric force. The perpendicular part of the right hand side in \Eqref{gceom} thus cancels, in accordance with the Lorentz transformation of the fields in the non-relativistic limit, $\vek{E}_\perp+\vek{U}\times\vek{B}=\vek{0}$. The parallel component of \Eqref{gceom} describes the uniform acceleration along the magnetic field, given by \Eqref{Uspardyn}. In the next section we will see how the fictitious forces are modified by weak variations in the electromagnetic fields.

\subsection{Gyro-average}

We know from \Secref{frame} that by choosing a frame of reference without any electric field, the particle motion only consists of gyration. The conceptual advantage of selecting such a frame of reference is that the motion relative to an inertial system is completely determined by \Eqref{gceom}, which describes the drifts and the acceleration of the guiding center frame relative to the inertial system. In the following, we use the knowledge of the solutions of \Eqref{eom} to show that one can separate the motion of the guiding center from the gyration motion without the need for a coordinate transformation. The underlying idea is to use that the gyration radius $\bxi$ and the gyration velocity $\vek{u}$ are periodic functions with period $2\pi/\omega$. This property provides the powerful result that 
the time average of the position vector $\vek{x}(t)$ over one gyration period gives the guiding center position,
\begin{equation} \label{gyroave}
\ave{\vek{x}} = \frac{\omega}{2\pi}\int_{t-\pi/\omega}^{t+\pi/\omega}\rmd t'\,\vek{x}(t') = \vek{X}(t) .
\end{equation}
Similarly, the gyro-average of the particle velocity gives the guiding center velocity
\begin{equation}
\ave{\vek{v}} = \vek{U}_\spar + \vek{U}_E = \vek{U}(t) .
\end{equation}
By applying the gyro-average on the equation of motion, it follows that the equation of motion for the guiding center drift is given by \Eqref{gceom}. On the other hand, the dynamics described in the guiding center system is given by \Eqref{quB}. This shows that the gyro-average can be used to define the guiding center variables and the gyration variables. For particle motion in slowly varying electromagnetic fields it will turn out that \eqref{gyroave} has an important significance in terms of being able to uniquely define the splitting between the guiding center motion and the gyration motion.

\subsection{Energy conservation}

In the case of constant fields, the electric field can be written in terms of the electrostatic potential as $\vek{E}=-\nabla\phi$. Multiplying the equation of motion \eqref{eomB} by $\vek{v}$ and using that $\phi$ is independent of time, we get an equation describing conservation of energy,
\begin{equation} \label{energy}
\ddt{}\left( \frac{1}{2}\,mv^2 + q\phi \right) = 0 .
\end{equation}
During its gyro-motion, the particle is alternately accelerated and decelerated by the $\vek{B}$-perpendicular components of the electric field. The resulting changes in the particle velocity $\vek{v}_\perp$ and the instantaneous center of gyration gives rise to a net drift across the electric and magnetic fields. Similarly, the kinetic energy associated with the parallel motion changes as the paritcle moves along the magnetic field into regions with smaller or larger potential energy.

In summary, the motion of a charged particle in constant and uniform electric and magnetic fields can be considered as a fast gyration around the guiding center position and a drift of the guiding center. The latter comprises the parallel motion described by \Eqref{Uspar} and a perpendicular electric drift given by \Eqref{ExB}. In the following section, this concept of a guiding center will be used to describe particle motion in weakly varying fields.

\section{Motion in weakly varying fields}\label{varying}

In the case of weakly varying electromagnetic fields $\vek{E}$ and $\vek{B}$, exact solutions to the equations of motion generally cannot be found.  However, approximate solutions can be obtained by a perturbation method based on expansion in the small parameter
\begin{equation}
\frac{\xi}{L} \sim \frac{\Omega}{\omega}  \sim \epsilon \ll 1,
\end{equation}
where $L$ is a characteristic length scale and $\Omega$ is a characteristic frequency for variations in the electromagnetic fields,
\begin{subequations} \label{ordering}
\begin{gather}
\frac{\abs{(\bxi\cdot\nabla)\vek{E}}}{\abs{\vek{E}}} \sim
\frac{\abs{(\bxi\cdot\nabla)\vek{B}}}{\abs{\vek{B}}} \sim \frac{\xi}{L} \sim \epsilon , \\
\frac{1}{\omega\abs{\vek{E}}}\abs{\frac{\p\vek{E}}{\p t}} \sim
\frac{1}{\omega\abs{\vek{B}}}\abs{\frac{\p\vek{B}}{\p t}} \sim \frac{\Omega}{\omega} \sim \epsilon .
\end{gather}
\end{subequations}
It is worth emphasizing that we here consider the case with spatial and temporal variations in the electric and magnetic fields of the same order of magnitude. In the following we discuss how this is achieved by use of a perturbation technique. It should be noted that in the limit of vanishing $\epsilon$ we recover the results for constant and uniform fields discussed in the previous section.

\subsection{Multi-scale expansion}\label{sec:multiscale}

From the discussion of particle motion in constant and uniform fields, we anticipate the existence of fast and slow temporal scales associated with the gyration and guiding centre motions, respectively. These are formally defined by
\begin{equation}
\tau = t ,
\qquad
T = \epsilon\,t .
\end{equation}
This means that the gyration motion occurs on a time scale that is $\order{1}$ and the slow drifts take place on a time scale that is $\order{1/\epsilon}$. In the following, these two temporal scales will be treated as distinct independent variables such that the partial time derivative transforms as
\begin{equation} \label{tgrad}
\frac{\p}{\p t} = \frac{\p}{\p\tau} + \epsilon\,\frac{\p}{\p T} .
\end{equation}

Similarly, the particle position can be written in terms of a fast rotating vector $\bxi$, corresponding to the gyration motion, and a slowly varying guiding centre position $\vek{X}$, formally defined by
\begin{equation}
\bxi=\vek{x} ,
\qquad
\vek{X} = \epsilon\,\vek{x} ,
\end{equation}
so that the del operator transforms as
\begin{equation} \label{xgrad}
\frac{\p}{\p\vek{x}} = \frac{\p}{\p\bxi} + \epsilon\,\frac{\p}{\p\vek{X}} .
\end{equation}
Accordingly, the particle velocity is decomposed into a fast rotating part, $\vek{u}$, and a slowly varying guiding centre drift, $\vek{U}$, by $\vek{v}=\vek{u}+\vek{U}$. The electromagnetic fields $\vek{E}$ and $\vek{B}$ are assumed to have slow temporal and weak spatial variations, that is, $\vek{E}=\vek{E}\left(\vek{X},T\right)$ and $\vek{B}=\vek{B}\left(\vek{X},T\right)$. As previously, we will use $\vek{b}=\vek{B}/B$ to denote the local unit vector along the magnetic field. It should be noted that the unit vectors $\vek{e}_j$ for $j=1,2,3$ all depend on the slow temporal scale $T$. The spatial coordinates $\vek{X}$ are dependent variables for the particle position but independent variables for the electromagnetic fields. The fast rotating gyration radius is a function of both the fast and the slow temporal scales, $\bxi=\bxi(\tau,T)$. The gyration velocity is given by $\vek{u}=\p\bxi/\p\tau$. The guiding center position only depends on the slow temporal scale, $\vek{X}(T)$.

The indeterminacy introduced by the multiple scales is lifted by the requirement that the solution must be periodic on the fast temporal scale. In particular, it follows that the gyro-average of the gyration vector must vanish,
\begin{equation} \label{gyroave1}
\ave{\bxi} = \frac{\omega}{2\pi}\int_{\tau-\pi/\omega}^{\tau+\pi/\omega}\rmd\tau'\,\bxi(\tau') = \mathbf{0} ,
\end{equation}
where $\omega=qB/m$ is the instantaneous frequency of gyration. It follows that the gyro-average of all fast temporal derivatives of $\bxi$ and $\vek{u}$ must also vanish. In particular, $\ave{\vek{u}}=\vek{0}$ and $\ave{\p\vek{u}/\p\tau}=\vek{0}$, which will be used repeatedly in the following.

\subsection{Guiding center frame}

\begin{figure}[t]
\centering
\includegraphics[width=8cm]{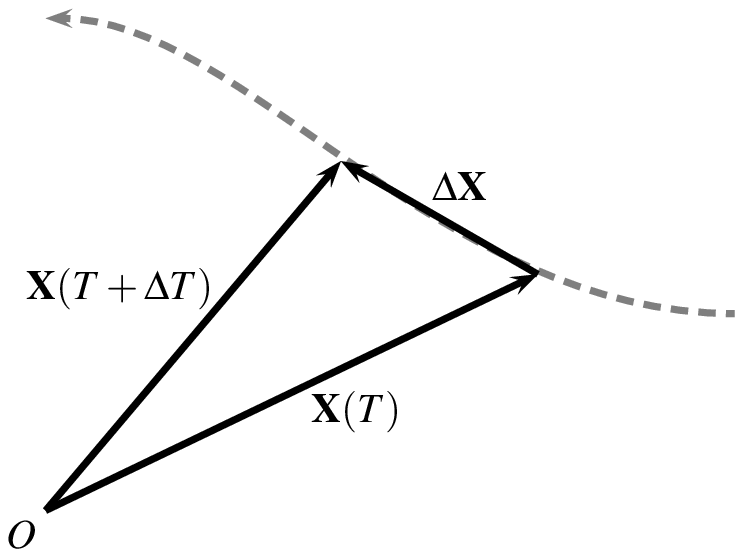}
\caption{Illustration of the guiding center trajectory (broken line) as observed in the inertial frame of reference with origin $O$. The separation $\Delta\vek{X}=\vek{X}(T+\Delta T)-\vek{X}(T)$ is the change in the guiding center position $\vek{X}$ over a time step $\Delta T$. The instantaneous guiding center velocity is given by $\vek{U}=\lim_{\Delta T\rightarrow 0}\Delta\vek{X}/\Delta T$.}
\label{fig:orbit}
\end{figure}

The total time rate of change of a quantity observed in the guiding center system is given by 
\begin{equation} \label{totaltgrad0}
\ddt{} = \pdt{} + \pdt{S}\frac{\partial}{\partial S} ,
\end{equation}
where $S$ is the distance along the guiding center trajectory. The first term on the right hand side of \Eqref{totaltgrad0} is the local rate of change at a fixed point in space, while the second term on the right hand side is the rate of change due to the effect of following the guiding center. According to the scale separation, the guiding center system is described by the slowly varying position $\vek{X}$ and the slow temporal scale $T$, thus $S=S(\vek{X}(T))$. It follows that the guiding center velocity is given by
\begin{equation}\label{velocityalongS}
\pdt{S} = \epsilon\,\pdT{S} ,
\end{equation} 
and the spatial rate of change $\partial/\partial S$ along the guiding center trajectory is according to the chain rule of differentiations given by
\begin{equation} \label{derivativalongS}
\frac{\partial}{\partial S} = \frac{\partial\vek{X}}{\partial S}\cdot\frac{\partial}{\partial\vek{X}} ,
\end{equation}
where $\p\vek{X}/\p S$ is the unit tangent vector along the trajectory as illustrated in \Figref{fig:orbit}. By substituting \Eqsref{tgrad}, \eqref{velocityalongS} and \eqref{derivativalongS} into the total time derivative in \Eqref{totaltgrad0} it follows that 
\begin{equation}\label{totaltgrad}
\ddt{} = \frac{\p}{\p\tau} + \epsilon\left(\frac{\p}{\p T} + \ddT{\vek{X}}\cdot\frac{\p}{\p\vek{X}} \right) ,
\end{equation}
where we have used that the guiding center velocity vector can be written as 
\begin{equation}
\vek{U} = \pdT{S}\frac{\p\vek{X}}{\p S} = \ddT{\vek{X}} .
\end{equation} 
Equation \eqref{totaltgrad} describes the time rate of change as observed in the guiding center frame of reference, which comprises explicit temporal variation on fast and slow temporal scales as well as variation due to motion of the guiding center.

Introducing the above separation of scales into the momentum equation \eqref{eomB} gives
\begin{equation} \label{eomms}
m \left( \pdtau{} + \epsilon\,\pdT{} + \epsilon\,\vek{U}\cdot\frac{\p}{\p\vek{X}} \right) \left( \vek{u} + \vek{U} \right) = q\vek{E} + q\left( \vek{u} + \vek{U} \right)\times\vek{B} ,
\end{equation}
Taking the gyro-average of \Eqref{eomms} gives the evolution equation for the guiding center velocity,
\begin{equation} \label{eommsgc}
\epsilon m \left( \pdT{} + \vek{U}\cdot\frac{\p}{\p\vek{X}} \right) \vek{U} = q\ave{\vek{E}} + q\vek{U}\times\ave{\vek{B}} + q\ave{\vek{u}\times\vek{B}} .
\end{equation}
Subtracting this from \Eqsref{eomms} gives the equation for the particle velocity associated with the fast gyration motion,
\begin{equation} \label{eommsgm}
m \left( \frac{\p}{\p\tau} + \epsilon\,\pdT{} + \epsilon\vek{U}\cdot\frac{\p}{\p\vek{X}} \right) \vek{u} = q(\vek{E}-\ave{\vek{E}}) + q(\vek{u}+\vek{U)}\times\vek{B} - q\ave{(\vek{u}+\vek{U})\times\vek{B}} .
\end{equation}
These equations provide the starting point for analyzing the particle motion in the case of weakly varying fields.

\subsection{Expansion of dependent variables}

A power series expansion is made for the dependent variables on the form
\begin{equation} \label{expansion}
\bxi = \sum_{n=0}^\infty \epsilon^n\bxi_n ,
\qquad
\vek{X} = \sum_{n=0}^\infty \epsilon^n\vek{X}_n ,
\qquad
\vek{u} = \sum_{n=0}^\infty \epsilon^n\vek{u}_n ,
\qquad
\vek{U} = \sum_{n=0}^\infty \epsilon^n\vek{U}_n .
\end{equation}
Accordingly, we expand the electromagnetic fields, which in \Eqref{eomms} are to be evaluated at the particle position, around the zero'th order slowly varying guiding center position $\vek{X}_0$,
\begin{subequations}  \label{fieldexpan}
\begin{align}
\vek{E} & = \vek{E}|_{\vek{X}=\vek{X}_0} + \left.\epsilon\left( \bxi_0\cdot\frac{\p}{\p\vek{X}} \right)\vek{E}\right|_{\vek{X}=\vek{X}_0} + \order{\epsilon^2}, \label{efieldexpan}
\\
\vek{B} & = \vek{B}|_{\vek{X}=\vek{X}_0} + \left.\epsilon\left( \bxi_0\cdot\frac{\p}{\p\vek{X}} \right)\vek{B}\right|_{\vek{X}=\vek{X}_0} + \order{\epsilon^2} . \label{mfieldexpan} 
\end{align}
\end{subequations}
Here and in the following, the electromagnetic fields and their spatial derivatives are to be evaluated at the zero'th order guiding centre position $\vek{X}_0$. For simplicity of notation, this will not be written out explicitly. In the following calculations, we will only need these first order expansions of the electromagnetic fields.

By substituting the expansion into \Eqref{eomms} and collecting parts of the same order we obtain equations determining the zero'th order velocities $\vek{u}_0$ and $\vek{U}_0$ and their corrections to higher orders. Substituting the multi-scale expansion into the equations of motion gives the evolution on the fast and slow spatio-temporal scales,
\begin{equation}
\left[ \pdtau{} + \epsilon\,\pdT{} + \epsilon\left( \sum_{n=0}^\infty \epsilon^n\vek{U}_n \right) \cdot \frac{\p}{\p\vek{X}} \right] \left(\sum_{n=0}^\infty \epsilon^n\left(\vek{u}_n+\vek{U}_n\right)\right) = q\vek{E} + q\left[ \sum_{n=0}^\infty \epsilon^n\left(\vek{u}_n+\vek{U}_n\right) \right]\times\vek{B} ,
\end{equation}
where $\vek{E}$ and $\vek{B}$ are given by \Eqref{fieldexpan}. The particle position vectors are related to the velocities by $\vek{u}_n=\p\bxi_n/\p\tau$ and $\vek{U}_n=\rmd\vek{X}_n/\rmd T$ for each order $n$.

\subsection{Particle gyration and drift motion}

To zero'th order in $\epsilon$, the equation of motion reads
\begin{equation} \label{gyrationvelocity0}
\pdtau{\vek{u}_0} = \omega\vek{u}_0\times\vek{b} ,
\end{equation}
which we know from \Secref{motion} describes uniform gyration. From the gyro-averaged equation of motion it follows that the zero'th order guiding center velocity is given by
\begin{equation} \label{guidingvelocity0}
\vek{E} + \vek{U}_0\times\vek{B} = \mathbf{0} . 
\end{equation}
This shows that the zero'th order perpendicular velocity $\vek{U}_{0\perp}$ is the electric drift given by \Eqref{ExB}, while the zero'th order parallel velocity component $U_{0\spar}$ is not determined to this order in the expansion.

Equation \eqref{guidingvelocity0} implies that there is no zero'th order parallel electric field component, that is, $|\vek{E}_{\spar}|/|\vek{E}_{\perp}|\sim\order{\epsilon}$, so that the total electric field can be written as 
\begin{equation} \label{Epp}
\vek{E} = \vek{E}_{\perp} + \epsilon\vek{E}_\spar. 
\end{equation}
An order unity parallel electric field component would violate the assumption of scale separation. This follows from the requirement that the particle must travel a short distance along the magnetic field over one gyration period, $U_\spar/\omega$, compared to the scale length of the electromagnetic fields, $L$. For this to hold on the slow temporal scale, the increment in the parallel particle velocity during one gyration period, given by $\Delta U_\spar\sim qE_\spar/m\omega\sim E_\spar/B$, must be much smaller than the gyration velocity, which can be estimated by the electric drift, $\xi\omega\sim E_\perp/B$, as discussed in \Secref{initial}. From this it follows that $E_\spar/E_\perp\ll1$.

To first order in $\epsilon$, the guiding center velocity described by \Eqref{eommsgc} becomes
\begin{equation} \label{guidingvelocity1}
m\,\ddT{\vek{U}_0} = q\vek{E}_{\spar} + q\vek{U}_1\times\vek{B} + q\ave{\vek{u}_0\times\left( \bxi_0\cdot\frac{\p}{\p\vek{X}} \right)\vek{B}} ,
\end{equation}
where we have defined the zero'th order advective derivative by
\begin{equation}
\ddT{} = \pdT{} + \vek{U}_0\cdot\frac{\p}{\p\vek{X}} .
\end{equation}
The last term on the right hand side in \Eqref{guidingvelocity1} is simplified by using \Eqref{uoxb} and the relation
\begin{align} \label{mmf}
\frac{m}{qB}\ave{ (\bxi_0\times\vek{b}) \times \left( \bxi_0\cdot\frac{\p}{\p\vek{X}} \right) \vek{B} } & = \vek{b}\ave{ \bxi_0 \cdot\left( \bxi_0\cdot\frac{\p}{\p\vek{X}} \right)\vek{B} } -
\ave{ \bxi\left[ \vek{b}\cdot\left( \bxi_0\cdot\frac{\p}{\p\vek{X}} \right)\vek{B} \right] } \notag
\\
& = \vek{b}\ave{ \bxi_0\bxi_0 }:\frac{\p\vek{B}}{\p\vek{X}} -
\ave{ \bxi_0\bxi_0 }\cdot\frac{\p B}{\p\vek{X}} \notag
\\
& = - \frac{mu_0^2}{2B} \frac{\p B}{\p\vek{X}} ,
\end{align}
where we have used $\vek{b}\cdot(\bxi_0\cdot\p/\p\vek{X})\vek{b}=0$, $\ave{\bxi_0\bxi_0}=(\xi_0^2/2)(\tensor{I}-\vek{b}\vek{b})$ and $\tensor{I}:\p\vek{B}/\p\vek{X}=(\p/\p\vek{X})\cdot\vek{B}=0$ with $\tensor{I}$ as the identity tensor. The equation of motion \eqref{guidingvelocity1} can thus be written as 
\begin{equation} \label{guidingvelocity2}
m\,\ddT{\vek{U}_0} = q\vek{E}_{\spar} + q\vek{U}_1\times\vek{B} - \mu_0\,\frac{\p B}{\p\vek{X}} ,
\end{equation}
where $\mu_0=mu_0^2/2B$ is the zero'th order magnetic dipole moment. This equation determines the zero'th order parallel velocity and the first order perpendicular guiding centre drift. With reference to the discussion in \Secref{frame}, we observe the presence of fictitious forces, with the term on the left hand side, $-m\,\rmd\vek{U}_0/\rmd T$, commonly refered to as the \emph{inertia force}. The last term on the right hand side, $\mu_0\,\p B/\p\vek{X}$, is sometimes referred to as the \emph{magnetic moment force}, due to the current loop comprised by the gyrating particle as discussed in \Secref{moment}. In the following we discuss how these forces give rise to parallel acceleration and drift of the guiding center perpendicular to the magnetic field.

Taking the projection of \Eqref{guidingvelocity2} along $\vek{B}$ gives the evolution equation for the zeroth order parallel velocity,
\begin{equation} \label{vpareq}
m\,\ddT{U_{0\spar}} = qE_{\spar} - m\vek{b}\cdot\ddT{\vek{U}_E} - \bmu_0\cdot\frac{\p B}{\p\vek{X}} ,
\end{equation}
where the second and third term on the right hand side are the parallel part of the inertia and magnetic moment forces, respectively.  The former may be rewritten as $m\vek{U}_E\cdot\rmd\vek{b}/\rmd T$, revealing an apparent parallel particle acceleration simply due to a change in the direction of the magnetic field as observed in the reference frame of the guiding center.  The last term in \Eqref{guidingvelocity2} is the \emph{magnetic mirror force}, which may gives rise to reflection of charged particles in regions with a converging magnetic field.

By taking the cross product of the momentum equation \eqref{guidingvelocity2} with $\vek{b}$ we get the first order perpendicular guiding center drift,
\begin{equation}\label{guidingvelocity1perp}
\vek{U}_{1\perp} = \frac{m}{qB}\,\vek{b}\times\left( \ddT{\vek{U}_0} + \frac{\mu_0}{m}\,\pdX{B} \right) .
\end{equation}
This drift consist of two parts, an \emph{inertial drift} and a \emph{magnetic gradient drift}, given respectively by the first and second terms on the right hand side of \Eqref{guidingvelocity1perp}. It should be noted that these drifts depend on both the charge and the mass of the particle. The inertia drift can be written as 
\begin{equation}
\frac{m}{qB}\,\vek{b}\times\ddT{\vek{U}_0} = \frac{mU_{0\spar}}{qB}\,\vek{b}\times\left( \pdT{} + \vek{U}_E\cdot\pdX{} \right) \vek{b} + \frac{mU_{0\spar}^2}{qB}\,\vek{b}\times\left( \vek{b}\cdot\pdX{} \right)\vek{b} + \frac{m}{qB}\,\vek{b}\times\ddT{\vek{U}_E} ,
\end{equation}
where the second term on the right hand side is known as the \emph{magnetic curvature drift} and the last term is commonly known as the \emph{polarization drift}. 

Note that the first order parallel drift $\vek{U}_{1\spar}$ of the guiding center is undetermined to this order. It is determined by taking the $\vek{B}$-parallel projection of the second order momentum equation for the guiding center, which we have not written out explicitly. Qualitatively, the first order parallel motion is of the same type but a small correction to the zero'th order drift described by \Eqref{vpareq}. On the other hand, the first order perpendicular motion is fundamentally different from the zero'th order electric drift as they all depend on both the particle mass and charge.

\subsection{Energy conservation}\label{sec:energy}

The evolution equation for the kinetic energy of the particle is given by taking the dot product of the equation of motion \eqref{eomms} with the total velocity $\vek{u}+\vek{U}$,
\begin{equation}\label{energycons}
\left[ \pdtau{} + \epsilon\left( \pdT{} + \vek{U}\cdot\pdX{} \right) \right] \left[ \frac{1}{2}\,m\left( \vek{u} + \vek{U} \right)^2 \right] = q\left( \vek{u} + \vek{U} \right)\cdot\left( \vek{E}_{\perp} + \epsilon\vek{E}_{\spar} \right),
\end{equation}
where we have omitted to write out the expansion of the particle position and velocity. To zero'th order in $\epsilon$ the equation for the kinetic energy reads 
\begin{equation} \label{energy0}
\pdtau{}\left( \frac{1}{2}\,mu_0^2 \right) = q\vek{U}_0\cdot\vek{E}_{\perp} = 0 ,
\end{equation}
which is consistent with the lowest order momentum equation \eqref{gyrationvelocity0} for the gyration motion. To first order in $\epsilon$ the gyro-averaged energy equation is given by 
\begin{equation} \label{energy1}
\ddT{}\left( \frac{1}{2}\,mu_0^2 + \frac{1}{2}\,mU_{0\spar}^2 + \frac{1}{2}\,mU_E^2 \right) = q\vek{U}_0\cdot\vek{E}_{\spar } + q\vek{U}_1\cdot\vek{E}_{\perp} + q\ave{\vek{u}_0\cdot\left(\bxi_0\cdot\frac{\p}{\p\vek{X}}\right)\vek{E}_{\perp}} .
\end{equation}
The last term in the above equation can be written as
\begin{equation}
q\omega\ave{\bxi_0\times\vek{b}\cdot\left(\bxi_{0}\cdot\frac{\p}{\p\vek{X}}\right)\vek{E}_{\perp}} = - \frac{1}{2}\,q\omega\xi_0^2\vek{b}\cdot\left( \frac{\p}{\p\vek{X}}\times\vek{E} \right) = \mu_0\,\pdT{B} .
\end{equation}
where we have used the same average as in the derivation of \Eqref{mmf} and the lowest order Faraday's law. The energy equation can thus be written as
\begin{equation} \label{Uenergy}
\ddT{}\left( \frac{1}{2}\,mu_0^2 + \frac{1}{2}\,mU_{0\spar}^2 + \frac{1}{2}\,mU_E^2 \right) = q\vek{E}_{\spar}\cdot\vek{U}_{0\spar} + q\vek{E}_{\perp}\cdot\vek{U}_{1\perp} + \mu_0\,\pdT{B} .
\end{equation}
This equation shows that the lowest order kinetic energy of the particle can change in two different ways: by the work done on the guiding center by the electric force and by acceleration of the gyrating particle due to the electromotive force.

Multiplying \Eqref{vpareq} with $U_{0\spar}$ gives the change in the kinetic energy associated with the parallel particle motion,
\begin{equation} \label{Uparenergy}
\ddT{}\left( \frac{1}{2}\,mU_{0\spar}^2 \right) = \vek{U}_{0\spar}\cdot\left( q\vek{E}_{\spar} -  \mu_0\,\frac{\p B}{\p\vek{X}} - m\,\ddT{\vek{U}_E} \right) ,
\end{equation}
revealing the work done by the parallel components of the magnetic moment and polarization forces. Similarly, taking the scalar product of \Eqref{guidingvelocity1perp} with the perpendicular electric field $\vek{E}_\perp$ and rearranging terms gives the evolution of the kinetic energy associated with the electric drift,
\begin{equation} \label{UEenergy}
\ddT{}\left( \frac{1}{2}\,mU_E^2 \right) = \vek{U}_1\cdot q\vek{E}_{\perp} - \vek{U}_E\cdot\left( m\,\ddT{\vek{U}_{0\spar}} + \mu_0\,\frac{\p B}{\p\vek{X}} \right) .
\end{equation}
The second term on the right hand side can be rewritten as $m\vek{U}_{0\spar}\cdot\rmd\vek{U}_E/\rmd T$. Subtracting \Eqsref{Uparenergy} and \eqref{UEenergy} from \Eqref{Uenergy} gives the evolution of the energy associated with the gyration motion on the slow temporal scale,
\begin{equation} \label{uEnergy}
\ddT{}\left( \frac{1}{2}\,mu_0^2 \right) = \mu_0\,\ddT{B} .
\end{equation}
The time derivative on the right hand side includes the rate of change due to drift of the guiding center. Indeed, the particle cannot distinguish between spatial and temporal variations as it moves in the weakly varying fields.

\begin{figure}[t]
\centering
\includegraphics[width=12cm]{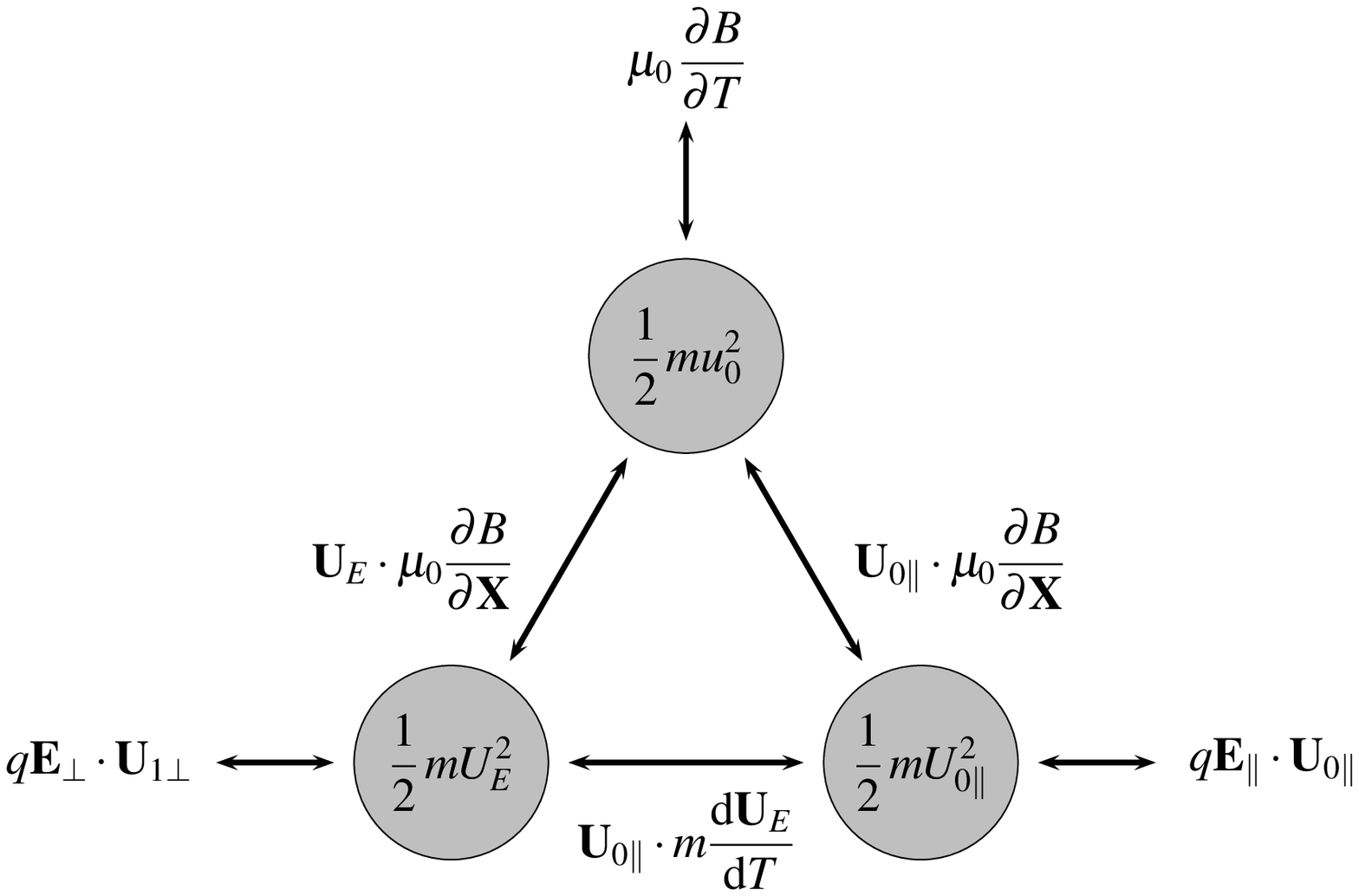}
\caption{Sources and transfer rates between energy associated with gyration, parallel motion and electric drift.}
\label{fig:energy}
\end{figure}

Since the magnetic field does not change on the fast temporal scale $\tau$, it follows from \eqref{energy0} that the magnetic moment is conserved on the fast scale,
\begin{equation}
\pdtau{\mu_0} = 0 .
\end{equation}
Moreover, from \Eqref{uEnergy} it follows that there is no change of the the zero'th order magnetic moment in the guiding center frame of reference,
\begin{equation}
\ddT{\mu_0} = \ddT{}\left( \frac{mu_0^2}{2B} \right) = 0 .
\end{equation}
Thus, the lowest order magnetic moment is a conserved quantity for the particle motion in the case of weakly varying electromagnetic fields. Magnetic moment conservation actually follows from invariance of the angular momentum of the particle in the guiding center frame of reference, which is given by
\begin{equation} \label{angmom}
\blambda = \bxi\times m\vek{u} .
\end{equation}
From the lowest order solution we find that the angular momentum is given by $\blambda_0=m\bxi_0\times\vek{u}_0=2m\bmu_0/q$. The temporal variation of the zero'th order angular momentum vector in the guiding center frame of reference is
\begin{equation}
\ddT{\blambda_0} = - \frac{2m}{q}\,\mu_{0}\,\ddT{\vek{b}} . 
\end{equation}
This implies that the angular momentum is to lowest order always aligned with the magnetic field, and the magnitude is conserved both on the fast and the slow temporal scales during the particle motion. Finally, it should be noted that the angular momentum invariance also implies that the lowest order magnetic flux through the gyration orbit is also conserved,
\begin{equation}
\Psi = \int \rmd\vek{S}\cdot\vek{B}
\end{equation}
which to zero'th order is given by $\Psi_0=\pi\xi_0^2B=2m\mu_0/q$. The conservation of magnetic flux for each particle corresponds to the familiar frozen-in field concept for magnetohydrodynamics.

\section{Conclusions}\label{sec:concl}

We have presented a new approach to the description of charged particle motion in weakly varying electromagnetic fields. The perturbation method presented here gives a direct and physically transparent description of the particle motion on fast and slow spatio-temporal scales, associated respectively with the particle gyration and field variations. This should be contrasted to the more complicated method of average first introduced Morozov and Solov'ev in Ref.~\onlinecite{ms} and used in most subsequent works on this problem.\cite{balescu,hm,hw} The method presented here can trivially be applied to the cases with no temporal or no spatial variations of the fields.

The lowest order particle motion comprises particle gyration, motion along the magnetic field and the electric drift across the magnetic field lines. By first considering the special case of constant and uniform fields, we introduced the magnetic field based coordinate system, the non-inertial guiding center frame of reference, and the magnetic dipole moment due to particle gyration. This leads to an intuitive generalization and scale separation for the case of weakly varying fields.

To next order in the expansion, numerous fictitious forces appear due to the weak spatio-temporal variations of the electromagnetic fields. A differential equation describes the particle motion along the magnetic field, including acceleration by the electric force as well as the magnetic mirror and polarization forces. The first order particle motion across magnetic field lines are given by derivatives of the electromagnetic fields, and include the magnetic gradient and curvature drifts and the well-known polarization drift.

We have given a detailed account of the fictitious forces due to acceleration of the guiding center frame of reference, the associated particle drifts and the energy transfer between the lowest order particle motion. Finally, invariance of the lowest order magnetic moment is shown to be the result of angular momentum conservation in the guiding center frame of reference. Interestingly, the assumption of a zero'th order perpendicular electric field corresponds to the magnetohydrodynamic ordering, implying conservation of magnetic flux through the particle gyro-orbit while the lowest order gyro-averaged equation of motion \eqref{guidingvelocity0} reduces to the ideal Ohm's law upon averaging over the velocity distribution of the charged particles in a plasma.


\end{document}